\documentclass{sig-alternate-2013}

\newfont{\mycrnotice}{ptmr8t at 7pt}
\newfont{\myconfname}{ptmri8t at 7pt}
\let\crnotice\mycrnotice%
\let\confname\myconfname%

\permission{Permission to make digital or hard copies of all or part of this work for personal or classroom use is granted without fee provided that copies are not made or distributed for profit or commercial advantage and that copies bear this notice and the full citation on the first page. Copyrights for components of this work owned by others than ACM must be honored. Abstracting with credit is permitted. To copy otherwise, or republish, to post on servers or to redistribute to lists, requires prior specific permission and/or a fee. Request permissions from permissions@acm.org.}
\conferenceinfo{SRIF'13,}{August 12, 2013, Hong Kong, China.}
\CopyrightYear{2013}
\crdata{978-1-4503-2181-5/13/08}
\begin{document}

\title{Software Defined Radio Implementation of Signaling Splitting in
Hyper-Cellular Network}
%\subtitle{[Extended Abstract]
%
% You need the command \numberofauthors to handle the 'placement
% and alignment' of the authors beneath the title.
%
% For aesthetic reasons, we recommend 'three authors at a time'
% i.e. three 'name/affiliation blocks' be placed beneath the title.
%
% NOTE: You are NOT restricted in how many 'rows' of
% "name/affiliations" may appear. We just ask that you restrict
% the number of 'columns' to three.
%
% Because of the available 'opening page real-estate'
% we ask you to refrain from putting more than six authors
% (two rows with three columns) beneath the article title.
% More than six makes the first-page appear very cluttered indeed.
%
% Use the \alignauthor commands to handle the names
% and affiliations for an 'aesthetic maximum' of six authors.
% Add names, affiliations, addresses for
% the seventh etc. author(s) as the argument for the
% \additionalauthors command.
% These 'additional authors' will be output/set for you
% without further effort on your part as the last section in
% the body of your article BEFORE References or any Appendices.

\numberofauthors{1} %  in this sample file, there are a *total*
% of EIGHT authors. SIX appear on the 'first-page' (for formatting
% reasons) and the remaining two appear in the \additionalauthors section.
%
\author{
% You can go ahead and credit any number of authors here,
% e.g. one 'row of three' or two rows (consisting of one row of three
% and a second row of one, two or three).
%
% The command \alignauthor (no curly braces needed) should
% precede each author name, affiliation/snail-mail address and
% e-mail address. Additionally, tag each line of
% affiliation/address with \affaddr, and tag the
% e-mail address with \email.
%
% 1st. author
\alignauthor
Tao Zhao, Pengkun Yang, Huimin Pan, Ruichen Deng, Sheng Zhou, Zhisheng Niu\\
       \affaddr{Tsinghua National Laboratory for Information Science and
       Technology}\\
       \affaddr{Department of Electronic Engineering}\\
       \affaddr{Tsinghua University, Beijing 100084, China}\\
       \email{t-zhao12@mails.tsinghua.edu.cn}
}
% There's nothing stopping you putting the seventh, eighth, etc.
% author on the opening page (as the 'third row') but we ask,
% for aesthetic reasons that you place these 'additional authors'
% in the \additional authors block, viz.
% Just remember to make sure that the TOTAL number of authors
% is the number that will appear on the first page PLUS the
% number that will appear in the \additionalauthors section.

\maketitle
\begin{abstract}

This paper presents the design and implementation of signaling splitting
scheme in hyper-cellular network on a software defined radio platform.
Hyper-cellular network is a novel architecture of future mobile
communication systems in which signaling and data are decoupled at the
air interface to mitigate the signaling overhead and allow energy
efficient operation of base stations.
On an open source software defined radio platform, OpenBTS, we investigate
the feasibility of signaling splitting for GSM protocol and implement a
novel system which can prove the proposed concept.
Standard GSM handsets can camp on the network with the help of
signaling base station, and data base station will be appointed to
handle phone calls on demand.
Our work initiates the systematic
approach to study hyper-cellular concept in real wireless environment
with both software and hardware implementations.

\end{abstract}

\category{C.2.1}{Computer-Communication Networks}%
{Network Architecture and Design}[Wireless communication]

\terms{Design, Experimentation}


\section{Introduction}

The mobile communication networks face the challenge of utilizing limited
spectrum and energy to meet the demand of rapid traffic growth.  In the
conventional cellular networks, each base station takes care of the coverage
as well as the traffic in its own cell. To cope with exponentially increasing
traffic demand of mobile Internet services~\cite{cisco2013vni},
more base stations are to be
deployed densely. It leads to large energy consumption, more inter-%
cell interference, and frequent handovers~\cite{cmri2011cran}.
Base station sleeping is proposed to increase the energy
efficiency, but can possibly generate wireless coverage holes which is
unacceptable in practical system.  Base station cooperation can be used to
mitigate inter-cell interference, but the scheduling of adjacent cells is hard
due to lack of global view.  Besides, frequent handovers increase
the signaling overhead, rendering the system less efficient.

The framework of hyper-cellular network aims at solving above problems by
separating the coverage of control channel and data channel~\cite{niu2012energy}.
In the hyper-cellular network, signaling splitting is realized with two types
of base stations: signaling base station (SBS) and data base station (DBS).
The SBS is responsible for the control coverage, while
DBSs take care of the data traffic demand and are deployed more densely.
Taking advantage of the space time variation of data traffic,
the DBSs can be dynamically switched on and off or
scheduled to provide coordinated transmission under the SBS's command.
By signaling splitting, hyper-cellular network can mitigate the signaling
overhead and make the system globally resource optimized and energy efficient.

Hyper-cellular network is a novel framework concept and the related study
focuses on the initial feasibility tests and simulations.
In the conceptual study based on LTE network~\cite{xu2013functionality}
Huawei researchers
propose to separate control and data by network functionalities.
Their numerical results show great reduction of energy consumption  of the separation scheme
compared to the tradition LTE networks.
However, no system implementation of the hyper-cellular architecture
is available in the literature to our knowledge.
Another question to the novel architecture is whether present wireless protocols
can be upgraded gradually to deploy the signaling splitting scheme.
The current cellular network is effectively a mixed system with
various protocols such as GSM, WCMDA, and LTE-A.
It will be a big incentive to attract the mobile operators and device vendors
to adopt the hyper-cellular architecture
if the current running system can be smoothly upgraded.
The paper takes GSM protocol as a case study and presents a working
demonstration of hyper-cellular network in which signaling splitting scheme is
implemented on a software defined radio platform.

Software defined radio originally means to move physical layer
functions from dedicated hardware
to programmable software in the radio communication system~\cite{ieee-std-dsa}.
Now its application has been extended to validate novel algorithms,
realize full wireless protocol stack, and implement state of the art wireless
systems~\cite{url:gnu-radio,url:openbts,kaszuba2011mimo,tan2011sora}.
The flexibility, rapid development and deployment,
ease of programming as well as low cost make software defined radio
a promising technology to power up new wireless communication systems.

% OpenBTS
In this paper we focus on the GSM air interface to study the realizability of
signal splitting of hyper-cellular network.  A working system is designed and
implemented on OpenBTS to prove the feasibility.
OpenBTS is a Unix application which implements the GSM air interface
with software defined radio technology~\cite{url:openbts}.
The public release is a free and open source software intended for
education, experimentation, and research projects.
It can run on a commodity PC and functions as a GSM base transceiver
station which allows standard handsets to register on its network,
make phone calls, and send text messages.
It uses Asterisk~\cite{url:asterisk}, a software PBX,
to switch traffic between different
users, therefore the necessity of
traditional core network equipment such as BSC and MSC is eliminated.
With a VoIP service provider, it can connect phone calls with various SIP
services and traditional cellular networks run by carriers.

The rest of the paper is
organized as follows.  Section 2 presents the system structure and base
station operation of hyper-cellular network.  Section 3 describes the GSM air
interface and propose our decoupling scheme of signaling and data.  We provide
the system implementation on OpenBTS in Section 4.  The conclusions are given
in Section 5.

\section{Base Station Operation in Hyper-Cellular Network}

\begin{figure}
\centering
\epsfig{file=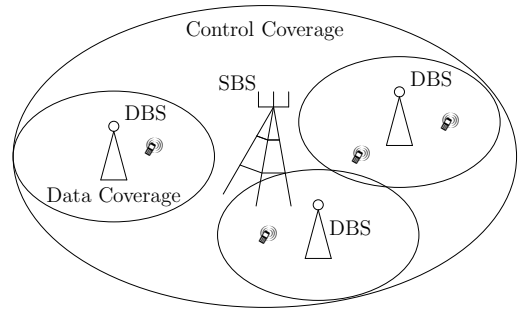, width=.38\textwidth}
\caption{The decoupling of control coverage and data coverage in
hyper-cellular network.}
\label{fig:overview}
\end{figure}

Hyper-cellular network emphasizes the separation of control channel and data
channel in the air interface and demands two types of base stations.
Since the network coverage is taken care of by signaling base stations,
data base stations can be switched off and go to sleep to save energy
when there is no traffic demand, and they can also be scheduled to provide
coordinated transmission to multiple users in multiple cells
to mitigate the inter-cell interference and improve the quality of service
(QoS). In the typical settings as in fig.~\ref{fig:overview}
the signaling base station has a larger coverage
where multiple data base stations reside and are scheduled by the signaling
base station.

The signaling base station is responsible for the control coverage.  When a mobile
handset is powered on and scans for existing network, the signaling base
station is supposed to check its identity and decide whether or not to allow
its camping. Once registered, the mobile user's network coverage will be handled
by the signaling base station. The context information is maintained between
the mobile user and the signaling base station, which does not involve the
nearby data base stations.

\begin{figure}
\centering
\epsfig{file=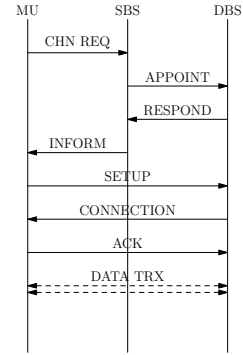, width=1.2in}
\caption{The information flow of mobile originated service.}
\label{fig:mos}
\end{figure}

The data base stations will be active only when there is data traffic service
associated. Fig.~\ref{fig:mos} shows the information flow between the mobile
user, the signaling base station and the data base stations in the mobile
originated service scenario.  When the mobile user initiates data service, it
will send the channel request to the signaling base station.  The signaling
base station will try to find one or more active data base stations to serve
the mobile user depending on the system information.  If the active data base
stations are all of high load but there are data base stations in sleep mode, the
signaling base station will try to wake up one or more sleeping data base
stations.  If there is no data base station available to provide the service,
the signaling base station will reject the mobile user's request.  Once the
appointed data base stations respond, the signaling base station will inform
the mobile user of the channel information of the granted access.  Then the
direct link between the mobile user and data base stations will be
established, followed by data traffic transmission.

\begin{figure}
\centering
\epsfig{file=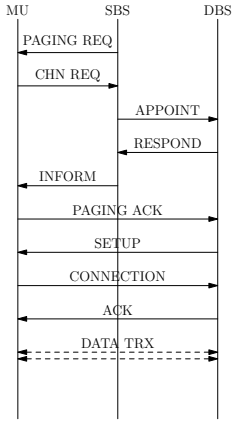, width=1.2in}
\caption{The information flow of mobile terminated service.}
\label{fig:mts}
\end{figure}

The case for mobile terminated service is shown in fig.~\ref{fig:mts} and
explained as follows.  When the mobile user has passive data service demand
such as an incoming call, it will receive the paging frames from the signaling
base station and then request necessary channel for the service.  The
signaling base station will try to appoint one or more available data base
stations. After receiving the response of data base stations, the signaling
base station will inform the mobile user of the assigned channel.  Then the
mobile user will acknowledge the paging to the data base stations.  Next the
appointed data base stations will establish the direct link to the mobile user
and provide the data traffic service.

\section{Signaling Splitting of GSM protocol}

The GSM air interface, namely Um, defines the layered structure, the frame
structure, and logical channels.  The Um layers roughly corresponds to the
bottom three layers of the OSI model.  The frequency band is divided by
FDMA into independent channels numbered by ARFCN.
TDMA is used to divide a frame into 8 time slots, each with the duration
0.577~ms. TDMA frames can be grouped to multiframes, and further to
superframes and hyperframes.
The standard defines multiple logical channels which are categorized into
traffic channels (TCH) and control channels (CCH).
The control channels can be divided into three groups:
broadcast channels (BCH) including BCCH, FCCH, and SCH;
common control channels (CCCH) consisting of PCH, NCH, RACH, and AGCH;
dedicated control channels (DCCH) which include SDCCH, SACCH, and
FACCH~\cite{eberspacher2009gsm}.

When the logical channels are mapped to the physical channel, only certain
combinations of logical channels are permitted according to the specification.
For example there are TCH+SACCH,  FCCH+SCH+BCCH+CCCH, and SDCCH+SACCH among
common combinations~\cite{halonen2003gsm}.
After the mapping the signaling in the control channel
and the data in the traffic channel are coupled together in the traditional
GSM cellular system.

Hyper-cellular network aims at decoupling signaling and data at the air
interface to enable dynamic network operation and soft resource matching.
When applying the signaling splitting concept into practical system,
we follow the principle of decoupling as much as possible while
ensuring smoothly upgradable. Thorough separation maximizes the benefits of
simplification and
specific optimization of two types of base stations, but might be impractical
to migrate existing system smoothly. The ability of upgrading smoothly is
highly attractive to network operators. No modification on the user
equipment side is appreciated as it brings no breakage to device vendors and
mobile users thus speeding up the migration.

Our signaling splitting scheme of GSM air interface is shown in
tab.~\ref{tab:functionality} and tab.~\ref{tab:logical}.
From the view of network functionalities, the signaling base station takes
care of synchronization, system information broadcasting, and paging,
while the data base station just handles data traffic services.
In the aspect of logical channels, BCH and CCCH are of the signaling base
station's responsibility, while TCH only resides at the data base station.
SACCH and FACCH are included on the data base station's side since they
carry the essential signaling during data transmission and the combination of
TCH+FACCH+SACCH is commonly used.
Note that the data base station has SDCCH as well as the signaling base station.
SDCCH is essentially a standalone logical channel and not tied to TCH,
but it is used in the SDCCH+SACCH combination to provide large SDCCH capacity
when the load is heavy. This consideration leads us to leave SDCCH also in the
data base station.

\begin{table}
\centering
\caption{Functionality separation of GSM air interface}
\label{tab:functionality}
\begin{tabular}{|c|c|c|} \hline
Functionality &Signaling BS&Data BS\\ \hline\hline
Synchronization & \checkmark & \\ \hline
Broadcasting & \checkmark & \\ \hline
Paging & \checkmark & \\ \hline
Data Traffic & & \checkmark \\ \hline
\end{tabular}
\end{table}

\begin{table}
\centering
\caption{Logical channel separation of GSM air interface}
\label{tab:logical}
\begin{tabular}{|c|c|c|} \hline
Logical channel &Signaling BS&Data BS\\ \hline\hline
BCH   & \checkmark & \\ \hline
CCCH  & \checkmark & \\ \hline
TCH   & & \checkmark \\ \hline
SACCH & & \checkmark \\ \hline
FACCH & & \checkmark \\ \hline
SDCCH & \checkmark & \checkmark \\ \hline
\end{tabular}
\end{table}

\section{Software Defined Radio Implementation}

We implement the proposed signaling splitting scheme of hyper-cellular network
concept on the software defined radio platform OpenBTS.  In the OpenBTS
application physical layer (L1) and data link layer (L2, LAPDm) are consistent
with the standard GSM Um interface specification, but the network layer (L3)
differs in several aspects: Radio Resource (RR) is terminated locally in
OpenBTS; Mobility Management (MM) and Call Control (CC) are translated to SIP
transactions and handed to external PBX~\cite{openbts-manual}.  Besides,
OpenBTS uses sqlite3 databases to store user information.  A
separate application named sipauthserve uses the databases to authenticate
mobile users.  This way traditional core network equipment is
avoided, leading to a flattened and lightweight network infrastructure.

Based on the source code of the OpenBTS application
we create the two types of base stations in hyper-cellular network.
The separation of network functionality and GSM logical channels
is applied to implement the signaling splitting scheme.
After the separation
only the signaling base station is capable of broadcasting system information,
discovering and authenticating mobile users.  The data base station keeps
silent until receiving the appointment from the signaling base station.
Then it will take care of the mobile user's data transmission.
The signaling message between the signaling base station and the data base
station is exchanged by UDP sockets in current implementation.

\begin{figure}
\centering
\epsfig{file=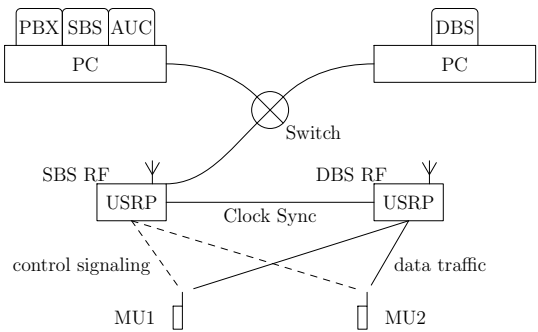, width=.4\textwidth}
\caption{The experiment scenario.}
\label{fig:exp}
\end{figure}

Figure~\ref{fig:exp} depicts our experiment scenario.
In the setup, two OpenBTS instances run on two Dell commodity desktop PCs.
The one with
Intel Core i7-2600 CPU serves as the signaling base station and
the other with
Intel Core2 Duo  E6550 CPU runs the affiliated data base station instance.
The two base stations use the same cell color code but different ARFCNs.
USRP N210s from Ettus Research~\cite{url:ettus} are used as
the radio front end. N210 is connected to the host PC by the Gigabit Ethernet
interface.
The on board
oscillator runs at the frequency 100~MHz, with the frequency stability of
2.5~ppm.  It can work fine with OpenBTS's transceiver with resampling enabled
at the host PC. We use WBX daughter boards which can operate in
the 900~MHz band. The antennas are of the type VERT900.
The two PCs and the master N210 are connected to one Gigabit Ethernet switch.
Two N210s are interconnected in shared Ethernet mode
with one MIMO cable which synchronizes the clock and timing between them
and enables data transmission between the slave N210 and the host PC.
Note that the asterisk instance and the sipauthserve
instance are running only on the signaling base station side.
They are in charge of switching and authentication respectively.
We use two Nokia 1000 GSM phones in the experiments.

In the tests we monitor the base stations' activities when mobile users camp on the
network and make phone calls to each other.
The experiment results show that standard GSM
handsets can camp on the network after powered on. The signaling information
is exchanged between the mobile handset and the signaling base station, while
the data base station does nothing but waits for the task appointment.
Phone calls can be connected between standard handsets.  The channel request
and paging are between the handset and the signaling base station. The channel
appointment and response are done between the two types of base stations.
Direct connection is then established between the data base station and the
mobile user for the connected phone call.  The results validate our signaling
splitting scheme of GSM air interface.  To our knowledge this is the
first working system in the real wireless environment to prove the
hyper-cellular network concept.

\section{Conclusions}

The hyper-cellular architecture is a redesign of the air interface in mobile
communication networks. It enables dynamic base station operation to
make the system globally resource optimized and  energy efficient
with the method of signaling
splitting. This  paper focuses on the GSM protocol and presents a feasible
way to implement signaling splitting scheme upon it. We implement the proposed
scheme on the software defined radio platform OpenBTS and create a prototype
system capable of discovering standard GSM handsets and connecting phone calls
in the real wireless environment. Our work demonstrates the hyper-cellular network
concept  in practical system and it can serve as the basis to bring
hyper-cellular network study into the system implementation field. Future
work will consider base station cooperation, evaluate the energy
efficiency, and investigate the signaling splitting scheme in more
advanced wireless protocols.

\section{Acknowledgments}

This work is sponsored in part by the National Basic Research Program of
China (2012CB316001), the Nature Science Foundation of China
(61201191, 60925002, 61021001), and Hitachi Ltd.

%
% The following two commands are all you need in the
% initial runs of your .tex file to
% produce the bibliography for the citations in your paper.
\bibliographystyle{abbrv}
%\bibliography{wireless}  % sigproc.bib is the name of the Bibliography in this case
% You must have a proper ".bib" file
%  and remember to run:
% latex bibtex latex latex
% to resolve all references
%
% ACM needs 'a single self-contained file'!
%
%APPENDICES are optional
%\balancecolumns
%\appendix
%Appendix A
%\section{Headings in Appendices}

% That's all folks!
\end{document}